\documentclass{ws-ijmpb}

\begin{document}

\markboth{Y.Hashizume and M.Suzuki}
{New Random Ordered Phase in Isotropic Models with Many-body Interactions}

\title{NEW RANDOM ORDERED PHASE IN ISOTROPIC MODELS \\WITH MANY-BODY INTERACTIONS}

\author{YOICHIRO HASHIZUME$^{*}$ and MASUO SUZUKI$^\dagger$}

\address{Depertment of Pure and Applied Physics, Graduate School of Science, Tokyo University of Science, 1-3, Kagurazaka, Shinjuku-ku, Tokyo 162-0825, Japan\\
$^{*}$ 1207707@ed.kagu.tus.ac.jp\\
$^\dagger $ msuzuki@rs.kagu.tus.ac.jp}
\maketitle

\begin{abstract}
 In this study, we have found a new random ordered phase in isotropic models with many-body interactions. 
Spin correlations between neighboring planes are rigorously shown to form a long-range order, namely coplanar order, using a unitary transformation, and the phase transition of this new order has been analyzed on the bases of the mean-field theory and correlation identities.
In the systems with regular 4-body interactions, the transition temperature $T_{\text{c}}$ is obtained as $T_{\text{c}}=(z-2)J/k_{\text{B}}$, and the field conjugate to this new order parameter is found to be $H^2$. In contrast, the corresponding physical quantities in the systems with random 4-body interactions are given by $T_{\text{c}}=\sqrt{z-2}J/k_{\text{B}}$ and $H^4$, respectively.
Scaling forms of order parameters for regular or random 4-body interactions are expressed by the same scaling functions in the systems with regular or random 2-body interactions, respectively.
Furthermore, we have obtained the nonlinear susceptibilities in the regular and random systems, where the coefficient $\chi_{\text{nl}}$ of $H^3$ in the magnetization shows positive divergence in the regular model, while the coefficient $\chi_{7}$ of $H^7$ in the magnetization shows negative divergence in the random model.
\end{abstract}

\keywords{Many-body interaction; coplanar correlations; random ordered phase}

\section{Introduction}
Recently, investigations of systems with many-body interactions are attractive even in the field of statistical mechanics of information, i.e., the learning theories using higher-order Boltzmann machines\cite{1}. 
Besides, some insulators have been known as materials composed mainly of 4-body interactions\cite{2,3,4}, as shown in Eq.(\ref{ham0}).
Study of phase transitions of systems with many-body interactions may be expected to give some useful hints for solving the above information theoretic problems.
Wegner\cite{5} concluded that spin systems only with 4-body interactions have no spin ordered phase on the basis of the dual transformation method in 1971.
However, one of the present authors (M.S.) solved rigorously this 4-body spin model with extremely anisotropic interactions ($J_{z}=0$) in three dimensions using the $\sigma$-$\tau$ transformation\cite{6,7}.
By this study of the extremely anisotropic model (we may call it no-ceiling model (fuki-nuke model in Japanese)) , Suzuki\cite{6} found the following interesting results, that is, (i) there exist an infinite number of degeneracies in the ground states, (ii) no spontaneous magnetization appears for all temperatures, (iii) spin correlations between neighboring planes (namely coplanar spin correlations) form a long-range order below the transition point $T_{\text{c}}$, (iv) the specific heat diverges at $T=T_{\text{c}}$.

Savvidy et.al.\cite{8,9,10} studied recently the same anisotropic 4-body spin model from a new view point of information theory and solved this model independently from Suzuki\cite{6} using the same $\sigma $-$\tau $ transformation.
Castelnovo et.al.\cite{11} studied also the same models from an information theoretic interest on glass transition, and they conjectured that similar orders may appear in the isotropic models.
It is not yet understood how the phase transition occurs and what configurations appear in the ordered phase of this isotropic model.

In order to analyze the new random ordered phase, we consider here the following model
\begin{equation}
\mathcal{H}=-\sum_{\text{plaquettes}}J_{ijkl}S_i S_j S_k S_l - \mu_{\text{B}}H\sum_{i}S_i, \label{ham0}
\end{equation}
only with 4-body interactions $\{J_{ijkl}\}$.
In section 2, spin correlations between neighboring planes are rigorously shown to form a long-range order using the unitary transformation, and the phase transitions are analyzed on the basis of the mean-field theory and correlation identities.
In the systems with {\itshape{regular}} 4-body interactions, the transition temperature $T_{\text{c}}$ is given by $T_{\text{c}}=(z-2)J/k_{\text{B}}$. In contrast, it is given by $T_{\text{c}}=\sqrt{z-2}J/k_{\text{B}}$ in the systems with {\itshape{random}} 4-body interactions.
Scaling forms of order parameters for regular or random 4-body interactions are expressed by the same scaling functions in the systems with regular or random 2-body interactions, respectively, as given in section 3.
In section 4, we derived the nonlinear susceptibilities, where the coefficient $\chi_{\text{nl}}$ of $H^3$ has a positive divergence in the regular model, while the coefficient $\chi_7$ of $H^7$ has a negative divergence in the random model.
In conclusion, it is understood how the coplanar spin correlations form a random ordered phase, and how the phase transitions occur in systems with many-body interactions.

\section{Degeneracy and coplanar orders}
As was mentioned in the previous section, the "no-ceiling model" with 4-body interaction was solved exactly by one of the present authors (M.S.)\cite{6} and was found to show very interesting behaviors in the low temperature phase below the transition point $T_{\text{c},0}$ (which agrees with that of the 2d square Ising model).
Fortunately, this is now found to play a crucial role in studying our isotropic model (\ref{ham0}). 
First, the transition point $T_{\text{c}}$ of this model is easily concluded to be higher than $T_{\text{c},0}$, using Griffiths inequalities\cite{12}.
Furthermore, the peculiar behaviors of the "no-ceiling model" are also proved to be preserved in our isotropic model by introducing an infinite number of unitary transformations, as shown below.

\begin{figure}[bt]
\centerline{\psfig{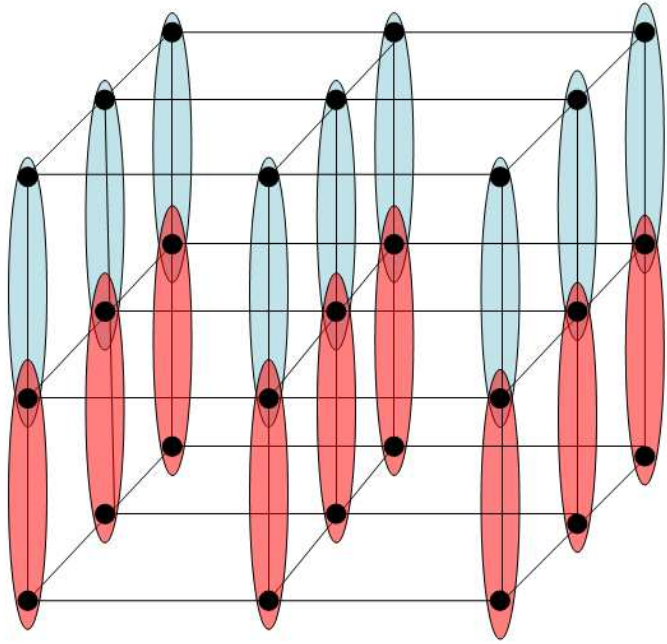}}
\vspace*{8pt}
\caption{This picture expresses the relationship of spin-pairs in the ground state, where the same color means the same relation (i.e. {(+,+)(-,-)} or {(+,-)(-,+)}). One can prove in any dimensions that similar configurations appear in other directions. }
\label{fig1}
\end{figure}

In general, using the Pauli operators $\{\sigma ^z_{(i_1,i_2,\dots ,i_d)}\}$ on the grid points $(i_1,i_2,\dots ,i_d)$ defined in a d-dimensional space including orthogonal coordinate axes $x_1,x_2,\cdots $, and $x_d$, the Hamiltonian with 4-body interactions $\{J^{lm}_{i_1,\cdots ,i_d}\}$ is given by
\begin{align}
\mathcal{H}=-\sum_{(l,m)}\sum_{(i_1,\cdots ,i_d)}J^{lm}_{i_1,\cdots ,i_d}&\sigma ^z_{(i_1,\dots ,i_l,\dots ,i_m,\dots ,i_d)}\sigma ^z_{(i_1,\dots ,i_l,\dots ,i_m+1,\dots ,i_d)} \notag
\\
\times &\sigma ^z_{(i_1,\dots ,i_l+1,\dots ,i_m,\dots ,i_d)}\sigma ^z_{(i_1,\dots ,i_l+1,\dots ,i_m+1,\dots ,i_d)}, \label{hamor}
\end{align}
where the natural numbers $l,m$, and $k$ are within the range $[1,d]$ and the natural numbers $\{i_k\}$ are within the range $[1,L]$ for the system-size $L$.
We introduce the following unitary transformations
\begin{equation}
\mathcal{U}_k\equiv \prod_{(i_1,\cdots ,\hat{i_k},\cdots ,i_d)}\sigma ^x_{(i_1,\cdots ,i_k^0 ,\cdots ,i_d)},
\end{equation}
which inverses all spins in the plane $x_k=i_k^0$.
The Hamiltonian $\mathcal{H}$ is invariant under any unitary transformation $\mathcal{U}_k$, that is, the Hamiltonian commutes all the unitary operators $\mathcal{U}_k$; $\mathcal{U}_k\mathcal{H}\mathcal{U}^{-1}_k=\mathcal{H}$.
Consequently, each order can appear independently in each plane.
Therefore, there appear an infinite number of ordered phases in the spatially isotropic model, similarly to the spatially anisotropic model\cite{7}.
Especially, the number of degeneracy of the ground states corresponds to the number of possible numbers of the unitary transformations. Clearly, there exist $d \times L$ combinations of $x_k$ and $i_k^0$ as give possible unitary transformations. Thus, the ground state has $2^{Ld}(\ll 2^{L^d})$ degeneracies.
Hence, it is shown that there exist the ordered states shown in Fig. \ref{fig1} in the region $T<T_{\text{c}}$, which appear independently in each plane.

Then, we prove that similar configurations appear in other directions.
The order parameters $\eta_{i_k}^{x_k}\equiv \langle S_{(i_1,\cdots ,i_k,\cdots ,i_d)}S_{(i_1,\cdots ,i_k+1,\cdots ,i_d)}\rangle$ ifor any $(i_1,\cdots ,\hat{i_k},\cdots ,i_d)$j are reexpressed in the operator forms $\eta_{i_k}^{x_k}\equiv \langle \sigma ^z_{(i_1,\cdots ,i_k,\cdots ,i_d)}\sigma ^z_{(i_1,\cdots ,i_k+1,\cdots ,i_d)}\rangle$.
The spin correlation
\begin{equation}
\eta_{i_k}^{x_k}\equiv \langle \sigma ^z_{(i_1,\cdots ,i_l^0,\cdots ,i_k,\cdots ,i_d)}\sigma ^z_{(i_1,\cdots ,i_l^0,\cdots ,i_k+1,\cdots ,i_d)}\rangle
\end{equation}
for $i_l=i_l^0$ is transformed into
\begin{equation}
\tilde{\eta}_{i_k}^{x_k} \equiv \mathcal{U}_l\eta_{i_k}^{x_k}\mathcal{U}^{-1}_l=\langle (-\sigma ^z_{(i_1,\cdots ,i_l^0,\cdots ,i_k,\cdots ,i_d)})(-\sigma ^z_{(i_1,\cdots ,i_l^0,\cdots ,i_k+1,\cdots ,i_d)})\rangle =\eta_{i_k}^{x_k}
\end{equation}
under the unitary transformation
\begin{equation}
\mathcal{U}_l=\prod_{(i_1,\cdots ,\hat{i_l},\cdots ,i_d)}\sigma ^x_{(i_1,\cdots ,i_l^0,\cdots ,i_d)},
\end{equation}
which inverses all spins in a plane $x_l=i_l^0$ for $l\not = k$.
The relation $\tilde{\eta}_{i_k}^{x_k}=\eta_{i_k}^{x_k}$ is proved trivially for $x_l=i_l\not=i_l^0$, as well.
Thus, the relationship 
\begin{equation}
\tilde{\eta}_{i_k}^{x_k}=\eta_{i_k}^{x_k}
\end{equation}
is proved for any $i_l$.
On the other hand, it is understood that $\eta_{i_l^0}^{x_l}$ can take each value $\pm |\eta_{i_l^0}^{x_l}|$ independently of $\eta_{i_k}^{x_k}$ since the coplanar orders perpendicular to $x_l$ axis $\eta_{i_l}^{x_l}=\langle S_{(i_1,\cdots ,i_l,\cdots ,i_d)}S_{(i_1,\cdots ,i_l+1,\cdots ,i_d)}\rangle$ (for any $(i_1,\cdots ,\hat{i_l},\cdots ,i_d)$) meet the conditions
\begin{equation}
\tilde{\eta}_{i_l^0}^{x_l}=\mathcal{U}_l\eta_{i_l^0}^{x_l}\mathcal{U}^{-1}_l=\langle (-\sigma ^z_{(i_1,\cdots ,i_l^0,\cdots ,i_d)})(\sigma ^z_{(i_1,\cdots ,i_l^0+1,\cdots ,i_d)})\rangle =-\eta_{i_l^0}^{x_l},
\end{equation}
and
\begin{equation}
\tilde{\eta}_{i_l\not= i_l^0}^{x_l}=\eta_{i_l\not= i_l^0}^{x_l}.
\end{equation}
All the coplanar order specified by $\{\eta_{i_k}^{x_k}\}$ appears at the same temperature (seen in sec.3).
Thus, similar configurations shown in Fig.\ref{fig1} are proved to appear in other directions.

\section{Analysis of order parameters}
\begin{figure}[bt]
\centerline{\psfig{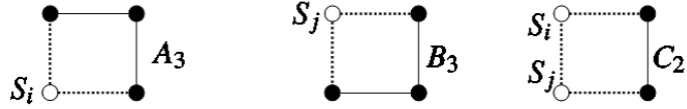}}
\vspace*{8pt}
\caption{Spins interacting $S_i,S_j$ directly under the 4-body interactions. The black filled sites mean the spins in a plaquette. Notation $A_3$ and $B_3$ express the summation of $(z-2)$ productions of three spins, while $C_2$ expresses the summation of $(z-2)$ productions of two spins.}
\label{fig2}
\end{figure}

In this section, we analyze the phase transitions of the isotropic systems with 4-body interactions using the correlation identities\cite{13,14}.
In order to apply these identities to our model, the Hamiltonian (\ref{ham0}) is decomposed into the following two parts
\begin{equation}
\mathcal{H}=\mathcal{H}_{ij}+\mathcal{H'},
\end{equation}
where the partial Hamiltonian $\mathcal{H'}$ does not include the nearest neighboring spins $S_i$ and $S_j$.
Then the partial Hamiltonian $\mathcal{H}_{ij}$ including the spins $S_i$ and $S_j$ is given in the form
\begin{equation}
\mathcal{H}_{ij}=-J(A_3 S_i+B_3 S_j+C_2 S_i S_j)-\mu_{\text{B}}H(S_i+S_j).\label{ham1}
\end{equation}
Here, $A_3$ includes all products of three spins interacting directly with the spin $S_i$. Similarly, $B_3$ includes all products of three spins interacting directly with the spin $S_j$. The term $C_2$, which is the most important term to analyze the coplanar order parameter, includes $(z-2)$ products of two spins interacting directly with the spins $S_i$ and $S_j$ (Fig. \ref{fig2}), where $z$ denotes the number of nearest neighbor spins.

According to the correlation identities\cite{13,14}, the spin correlation $\langle S_i S_j \rangle$ is derived easily as follows
\begin{align}
\langle S_i S_j \rangle&=\langle \langle S_i S_j \rangle _{\mathcal{H}_{ij}}\rangle\notag
\\
&=\left\langle\frac{\tanh K C_2 +\tanh (K A_3 +h)\tanh(K B_3 +h)}{1+\tanh K C_2 \tanh (K A_3 +h)\tanh(K B_3 +h)} \right\rangle\notag
\\
&\simeq (z-2)K\langle S_k S_l \rangle +h^2, \label{eq:app}
\end{align}
using decoupling approximations and relationships $\langle A_3\rangle=\langle B_3\rangle=\langle A_3B_3\rangle=0$.
These decoupling approximations are equivalent of mean-field approximations.
The parameters $h$ and $K$ are defined by $h=\mu_{\text{B}}H/k_{\text{B}}T$ and $K=J/k_{\text{B}}T$, respectively.
The self-consistency $\eta\equiv\langle S_i S_j\rangle=\langle S_k S_l\rangle $, which requests the translation symmetry of coplanar spin-correlations, yields the order parameter $\eta$ for small $H$ and for $T>T_{\text{c}}$ as follows
\begin{equation}
\eta=\langle S_i S_j \rangle =\frac{\mu_{\text{B}}^2}{k_{\text{B}}T}\frac{H^2}{k_{\text{B}}T-(z-2)J}.
\end{equation}
Therefore we obtain the transition temperature $T_{\text{c}}=(z-2)J/k_{\text{B}}$ and the external field $H^2$ conjugate to the order parameter $\eta$.

\begin{figure}[bt]
\centerline{\psfig{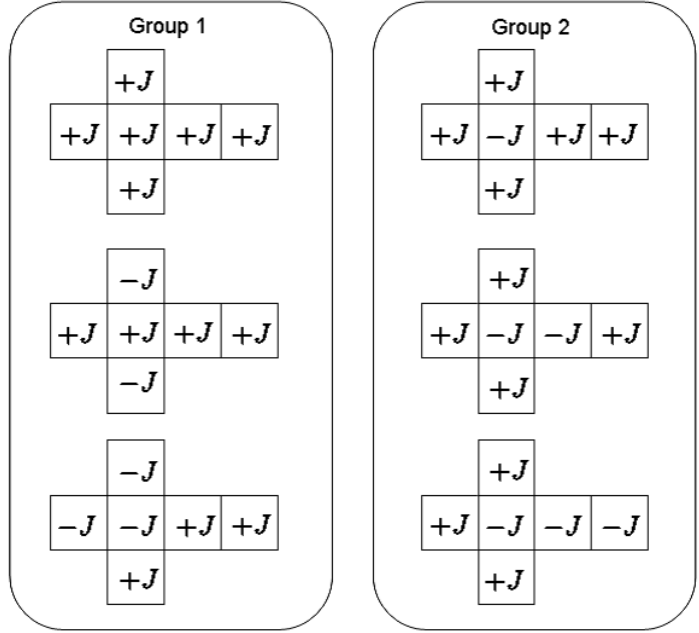}}
\vspace*{8pt}
\caption{Configurations of random 4-body interactions on a three dimensional unit cell. This figure shows all the possible configurations of random 4-body interactions $\pm J$ on the developed figure of a three-dimensional unit cell. All unit cells are categorized according to the frustration, that is, non-frustrated unit cells are categorized into "Group 1", while frustrated unit cells are categorized into "Group 2". Same configurations under the inversion of all $\{J_{ijkl}\}$ are omitted in this figure.}
\label{fig3}
\end{figure}

When the 4-body interactions $\{\pm J\}$ are distributed randomly, the number of frustrated unit cells is just the same as that of non-frustrated unit cells (Fig.\ref{fig3}).
Then, the order parameter $\zeta$ is defined using the double average $[\langle S_i S_j\rangle^2]$, where $[\cdots]$ expresses the random average for the distribution of $\{\pm J\}$.
Similarly to the regular model, the order parameter $\zeta$ is obtained as
\begin{equation}
\zeta \propto \frac{H^4}{k_{\text{B}}T-\sqrt{z-2}J}.
\end{equation}
Then we also obtain the transition temperature $T_{\text{c}}=\sqrt{z-2}J/k_{\text{B}}$ and the  external field $H^4$ conjugate to the order parameter$\zeta$ for random 4-body interactions.

The above results are listed in Table\ref{table1}.
\begin{table}[hpb] 
\tbl{The table of order parameters, critical temperatures, conjugate external fields and universality classes for four kinds of interactions. The critical temperatures of the random models are lower than those of regular models. The powers of conjugate external fields correspond to the number of spins included in the order parameters. The regular and the random models have their distinctive universality classes.}
{\begin{tabular}{lllcc} \Hline 
\\[-1.8ex] 
Interactions&order parameter&$k_{\text{B}}T_{\text{c}}$&field&universality class$(\gamma,1/\beta,\delta)$\\[0.8ex] 
\hline \\ [-1.8ex] 
regular 2-body(Ferro)\cite{15}&$m=\langle S_i \rangle$&$zJ$&$H$&$(1,2,3)$
\\
random 2-body(SG)\cite{7,16,17}&$q=[\langle S_i \rangle^2]$&$\sqrt{z}J$&$H^2$&$(1,1,2)$
\\
regular 4-body&$\eta=\langle S_i S_j \rangle$&$(z-2)J$&$H^2$&$(1,2,3)$
\\
random 4-body&$\zeta =[\langle S_i S_j \rangle^2]$&$\sqrt{z-2}J$&$H^4$&$(1,1,2)$
\\
metal-insulator transition\cite{18}&{}&{}&{}&$(1,1,2)$
\\
\Hline \\ [-1.8ex] 
\end{tabular}}
\label{table1}
\end{table}
There is no ergodic path\cite{19} by which the system can move from one ground state to another ground state only by the local correlation inversion in the systems with regular 4-body interactions, because regular models have no frustration in any unit cell.
For this reason, spin-correlations in the regular models need larger fluctuation for disordered states than in the random models.
Therefore the transition temperature of regular systems is higher than that of random systems.

Furthermore, scaling forms of the order parameters are also obtained near the transition point $(T\gtrless T_{\text{c}})$, by taking into account the nonlinear terms of equations of state derived by the correlation identities.
The equation of state of the order parameter $\eta$ in systems with regular 4-body interactions is obtained as $\eta^3+3 t \eta -3h^2=0$ for $t\equiv (T-T_{\text{c}})/T_{\text{c}} $. Then the scaling form of the order parameter $\eta$ is derived in the form
\begin{equation}
\eta=t^{1/2}f\left(\frac{h^2}{t^{3/2}} \right),
\end{equation}
where the scaling function $f(x)$ is obtained as the solution of the cubic equation $f^3+3f-3x=0$.
This function $f(x)$ is known as the scaling function of the ferromagnetic model constructed of regular 2-body interactions.
In addition to that, the power $2$ of $h^2$ in the scaling parameter corresponds to the number of spins included in the order parameter $\eta=\langle S_i S_j\rangle$.
In a similar way, the equation of state of the order parameter $\zeta$, namely $\zeta ^2+t \zeta -\frac{h^4}{2}=0$, leads to the scaling form of $\zeta$ as follows
\begin{equation}
\zeta=tf\left(\frac{h^4}{t^2} \right);\quad f(x)=\frac{1}{2}\left(\sqrt{1+2x}-1\right).
\end{equation}
Here, the scaling function $f(x)$ is the same as that of the spin-glass model constructed of random 2-body interactions\cite{16}. By the way, the power $4$ of $h^4$ in the scaling parameter corresponds to the number of spins included in the order parameter $\zeta=[\langle S_i S_j\rangle^2]$.

Additionally, the critical exponents $\beta,\gamma,\delta$ are obtained immediately from the equations of state (or the scaling forms of order parameters). 
Then the universality classes are expressed by the sets of the critical exponents $(\gamma,1/\beta,\delta)$.
In the present case, the universality class of our regular 4-body model is given by $(\gamma,1/\beta,\delta)=(1,2,3)$, while that of random 4-body model is given by $(\gamma,1/\beta,\delta)=(1,1,2)$.
These universality classes are also listed in Table\ref{table1}.
It is clearly understood that the universality class of each model depends on whether spin interactions are regular or random.
Especially, the universality class of the random 4-body model is the same as that of the spin-glass model (i.e.,the random 2-body model) which was obtained by one of the present authors (M.S.) in 1977\cite{7}.
The universality class of metal-insulator transitions belongs to this category, as was shown phenomenologically by March, Suzuki and Parrinello\cite{18}.

\section{Nonlinear susceptibility}
In this section, the singularity of the nonlinear magnetic susceptibility\cite{7} is analyzed near the transition point $T_{\text{c}}$, because it is useful in observing the coplanar random ordered states experimentally.
For the case of regular 4-body interactions, the partial Hamiltonian $\mathcal{H}_i$ including the spin $S_i$ is obtained as 
\begin{equation}
\mathcal{H}_i=-J A_3 S_i-\mu_{\text{B}}HS_i,
\end{equation}
using $A_3$ ($A_3$ is also used in eq.(\ref{ham1})).
According to the mean-field decoupling approximation
\begin{equation}
\langle S_j S_k S_l \rangle=\langle S_j \rangle\langle S_k S_l \rangle + \langle S_k \rangle\langle S_j S_l \rangle +\langle S_l \rangle\langle S_j S_k \rangle =2m\eta
\end{equation}
due to the fact that one of $\langle S_j S_k \rangle,\langle S_k S_l \rangle$ and $\langle S_j S_l \rangle$ is vanishing,
we derive the following equation of state
\begin{equation}
m=6(z-2)Km\eta +h
\end{equation}
from the correlation identity $\langle S_i \rangle =\langle \tanh(K A_3 +h) \rangle$ under the first-order approximation with respect to $m,\eta$ and $h$.
Since the relations $m\sim \chi_0 H$ and $\eta\sim H^2/(T-T_{\text{c}})$ hold near the transition point, the magnetization $m$ is obtained as follows
\begin{equation}
m=\chi_0 H + \chi_{\text{nl}} H^3+\cdots;\quad \chi_{\text{nl}}\propto +\frac{1}{T-T_{\text{c}}},
\end{equation}
which shows a positive divergence in the coefficient of $H^3$. 

Similarly to the regular model, the nonlinear magnetic susceptibility in the random model is obtained as follows:
\begin{equation}
m=\chi_0 H+\chi_3 H^3+\chi_5 H^5+\chi_7 H^7+\cdots;\quad \chi_{7}\propto -\frac{1}{T-T_{\text{c}}},
\end{equation}
which shows a negative divergence in the coefficient of $H^7$, while $\chi_3$ and $\chi_5$ are finite at $T=T_{\text{c}}$.

\section{Summary and Conclusion}
We have found a random ordered phase in the isotropic (even regular) Ising models with four-body interactions, as was conjectured by Castelnovo, Chamon and Sherrington\cite{11}.
Spin correlations between neighboring planes are rigorously shown to form a long-range order (coplanar order) using an infinite number of unitary transformations, and the detailed properties of this phase transition have been analyzed on the basis of the mean-field theory and correlation identities.

Our main results are the following:
(1) The order parameters of our models both for regular and random 4-body interactions are represented by the coplanar spin correlations.
(2) The external field conjugate to the order parameter $\eta=\langle S_i S_j\rangle$ is given by $H^2$ in the regular system. The external field conjugate to the order parameter $\zeta=[\langle S_i S_j\rangle^2]$ is obtained as $H^4$ in the random system.
(3) The scaling functions of the order parameters are the same as for 2-body interactions.
(4) We have clarified the singularity of nonlinear susceptibilities in the regular or random systems by studying the magnetization in each system, as in Ref.\refcite{7}.
The coefficient of the $H^3$ form in the nonlinear susceptibilities shows a positive divergence in the regular model, and the coefficient of the term $H^7$ , for the first, shows a negative divergence in the random model.
Finally we should remark that the nonlinear susceptibilities are crucially useful to study phase transitions experimentally or numerically on the systems with many body interactions.

\section*{Acknowledgements}
The authors would like to thank Professor D. Sherrington for informing Ref.\refcite{11} at Cairns in July, 2010.
One of the authors (Y.H.) thanks Professor X.Hu and Dr. M. Ohzeki for useful discussions.
The other author  (M.S.) is partially supported by the Grant-in-Aid for Scientific Research on Priority Area "Deepening and Expansion of Statistical Mechanical Informatics".

\appendix{Correlation function}
\begin{figure}[bt]
\centerline{\psfig{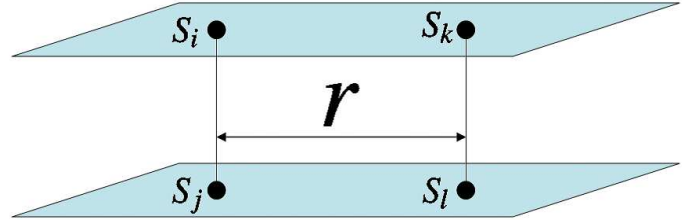}}
\vspace*{8pt}
\caption{Coplanar correlation of spin correlations. The spins $S_i,S_j,S_k$ and $S_l$ included in the correlation function $C(\vec{r})=\langle (S_i S_j)(S_k S_l)\rangle$ are located as shown.}
\label{fig4}
\end{figure}
Correlation functions between coplanar spins are derived from the correlation identities\cite{13,14}.
The correlation function $C(\vec{r})$ is defined by
\begin{equation}
C(\vec{r})=\langle (S_i S_j)(S_k S_l)\rangle,
\end{equation}
where the spins $S_i,S_j,S_k$ and $S_l$ are located as shown in Fig.\ref{fig4}.
The correlation function $C(\vec{r})$ for the Hamiltonian (\ref{ham1}) for $H=0$ is derived as follows
\begin{equation}
C(\vec{r})=\langle S_k S_l \langle S_i S_j \rangle_{\mathcal{H}_{ij}} \rangle = \langle S_k S_l \tanh KC_2 \rangle
\simeq K\sum_{i=1}^{z-2} \{ C(\vec{r}+a \vec{e}_{i})+C(\vec{r}-a \vec{e}_{i}) \},
\end{equation}
where $a$ denotes the lattice constant and $\vec{e}_{i}$ denotes the unit vector parallel to the $x_i$-axis.
In the second order-approximation of $C(\vec{r}\pm a \vec{e}_{i})$ with respect to the lattice constant $a$, we obtain the differential equation 
\begin{equation}
\left(\Delta -\frac{1-(z-2)K}{Ka^2}\right)C(\vec{r})=0;\quad \Delta\equiv \sum_{i=1}^{z-2}\frac{\partial^2 }{\partial x_i^2},\label{a3}
\end{equation}
which has the solution
\begin{equation}
C(\vec{r})\sim \frac{e^{-r/\xi }}{r}
\end{equation}
for $r\equiv |\vec{r}|$.
This solution $C(\vec{r})$ shows the Ornstein-Zernike type correlation function because of the mean-field approximations.
Furthermore, the correlation length $\xi$ is given by
\begin{equation}
\xi =\left[Ka^2/(1-(z-2)K)\right]^{1/2},
\end{equation}
as is easily seen from Eq.(\ref{a3}).

\end{document}